\documentclass[manuscript]{aastex}

\shorttitle{Collapsed Cores in Globular Clusters}
\shortauthors{Djorgovski et al.}

\begin{document}

%% LaTeX will automatically break titles if they run longer than
%% one line. However, you may use \\ to force a line break if
%% you desire.

\title{Signatures of black hole spin in galaxy evolution \\
    }

%% Use \author, \affil, and the \and command to format
%% author and affiliation information.
%% Note that \email has replaced the old \authoremail command
%% from AASTeX v4.0. You can use \email to mark an email address
%% anywhere in the paper, not just in the front matter.
%% As in the title, use \\ to force line breaks.

\author{D. Garofalo }
%\altaffilmark{1} }
\affil{Jet Propulsion Laboratory, California Institute of
Technology, Pasadena CA 91109 }
\email{David.A.Garofalo@jpl.nasa.gov}

%% Notice that each of these authors has alternate affiliations, which
%% are identified by the \altaffilmark after each name.  Specify alternate
%% affiliation information with \altaffiltext, with one command per each
%% affiliation.

%\altaffiltext{1}{Visiting Astronomer, Cerro Tololo Inter-American Observatory.
%CTIO is operated by AURA, Inc.\ under contract to the National Science
%Foundation.}
%\altaffiltext{2}{Society of Fellows, Harvard University.}
%\altaffiltext{3}{present address: Center for Astrophysics,
%    60 Garden Street, Cambridge, MA 02138}
%\altaffiltext{4}{Visiting Programmer, Space Telescope Science Institute}
%\altaffiltext{5}{Patron, Alonso's Bar and Grill}

%% Mark off your abstract in the ``abstract'' environment. In the manuscript
%% style, abstract will output a Received/Accepted line after the
%% title and affiliation information. No date will appear since the author
%% does not have this information. The dates will be filled in by the
%% editorial office after submission.

\begin{abstract}
We explore the connection between black hole spin and AGN power by
addressing the consequences underlying the assumption in the recent
literature that the gap region between accretion disks and black holes
is fundamental in producing strong, spin-dependent, horizon-threading
magnetic fields.  Under the additional assumption that jets and
outflows in AGN are produced by the Blandford-Znajek and
Blandford-Payne mechanisms, we show that maximum jet/outflow power is
achieved for accretion onto black holes having highly retrograde spin
parameter, an energetically excited yet unstable gravitomagnetic
configuration.

\end{abstract}

%% Keywords should appear after the \end{abstract} command. The uncommented
%% example has been keyed in ApJ style. See the instructions to authors
%% for the journal to which you are submitting your paper to determine
%% what keyword punctuation is appropriate.


%% From the front matter, we move on to the body of the paper.
%% In the first two sections, notice the use of the natbib \citep
%% and \citet commands to identify citations.  The citations are
%% tied to the reference list via symbolic KEYs. The KEY corresponds
%% to the KEY in the \bibitem in the reference list below. We have
%% chosen the first three characters of the first author's name plus
%% the last two numeral of the year of publication as our KEY for
%% each reference.

%% Authors who wish to have the most important objects in their paper
%% linked in the electronic edition to a data center may do so by tagging
%% their objects with \objectname{} or \object{}.  Each macro takes the
%% object name as its required argument. The optional, square-bracket 
%% argument should be used in cases where the data center identification
%% differs from what is to be printed in the paper.  The text appearing 
%% in curly braces is what will appear in print in the published paper. 
%% If the object name is recognized by the data centers, it will be linked
%% in the electronic edition to the object data available at the data centers  
%%
%% Note that for sources with brackets in their names, e.g. [WEG2004] 14h-090,
%% the brackets must be escaped with backslashes when used in the first
%% square-bracket argument, for instance, \object[\[WEG2004\] 14h-090]{90}).
%%  Otherwise, LaTeX will issue an error. 

\section{Introduction}

The paradigm that has emerged for the production of outflows from
active galactic nuclei (AGN) involves the presence of large scale
electromagnetic fields which are instrumental in their formation,
acceleration and collimation, many gravitational radii from the
central supermassive black hole (Nakamura et al, 2008; Meier et al,
2001; Blandford, 1976; Lovelace, 1976).  Two models have taken center
stage.  Blandford \& Payne (1982; henceforth BP) and extensions of
this model (Li et al, 1992 and Vlahakis \& Konigl, 2003) describe a
centrifugally driven outflow of gas originating in a cold accretion
disk as a solution to ideal MHD within the context of self-similarity.
If the angle between the poloidal component of the magnetic field and
the disk surface is less than 60 degrees, mass-loading of the magnetic
field lines occurs, leading to an inbalance between inward
gravitational and outward centrifugal forces, with gravity being
overwhelmed.  Unlike the BP mechanism which taps into the
gravitational potential energy of the accretion flow, the
Blandford-Znajek (1977; henceforth BZ) mechanism produces relativistic
jets from large scale magnetic fields threading the rotating event
horizon by extraction of black hole rotational energy.  The
flux-trapping model (Reynolds et al, 2006) is an attempt to understand
ways in which black hole accretion flows can overcome their diffusive
character (see also Bisnovatyi-Kogan \& Lovelace, 2007 and Rothstein
\& Lovelace, 2008) to produce strong magnetic field on the black hole
(see Bisnovatyi-Kogan \& Ruzmaikin, 1976, for earliest attempt to study
the accretion of large-scale ordered magnetic field on black
hole) indicating that if the flux-trapping behavior of the gap/plunge
region is valid, the BZ mechanism produces greatest power for black
hole spin of $a\approx-1$ (Garofalo, 2009).  Here we show that the
same is true for the BP mechanism.  This means that although the spin
dependence of BZ and BP power is different overall, they both peak for
near maximal retrograde black hole spin.  We motivate the idea that
'ordinary' astrophysical processes will tend to shift near maximal
retrograde black hole accretion systems toward more prograde spins
(i.e. accretion and/or spin energy extraction).  Once formed (e.g. in
galaxy mergers), such systems will evolve toward a state of lower
power output, which implies that the population density of near
maximal retrograde black hole accretion systems that produce outflows
and jets, is larger at the redshift of formation of the highly
retrograde accretion systems and naturally tends to drop, so that the
cosmological evolution of black hole spin is in the direction of
prograde spins.  In section 2 we describe the basic geometry of the
flux-trapping model.  In section 3 we discuss its implications for the
BP power and those of assuming that outflows and jets in AGN are all
due to either BZ, BP, or a combination of both mechanisms (Meier,
1999).  In section \ref{conclusion} we conclude.

\section{The model}

%% In a manner similar to \objectname authors can provide links to dataset
%% hosted at participating data centers via the \dataset{} command.  The
%% second curly bracket argument is printed in the text while the first
%% parentheses argument serves as the valid data set identifier.  Large
%% lists of data set are best provided in a table (see Table 3 for an example).
%% Valid data set identifiers should be obtained from the data center that
%% is currently hosting the data.
%%
%% Note that AASTeX interprets everything between the curly braces in the 
%% macro as regular text, so any special characters, e.g. "#" or "_," must be 
%% preceded by a backslash. Otherwise, you will get a LaTeX error when you 
%% compile your manuscript.  Special characters do not 
%% need to be escaped in the optional, square-bracket argument.

\begin{figure*}[h!]
\centerline{\includegraphics[angle=-90,scale=0.4]{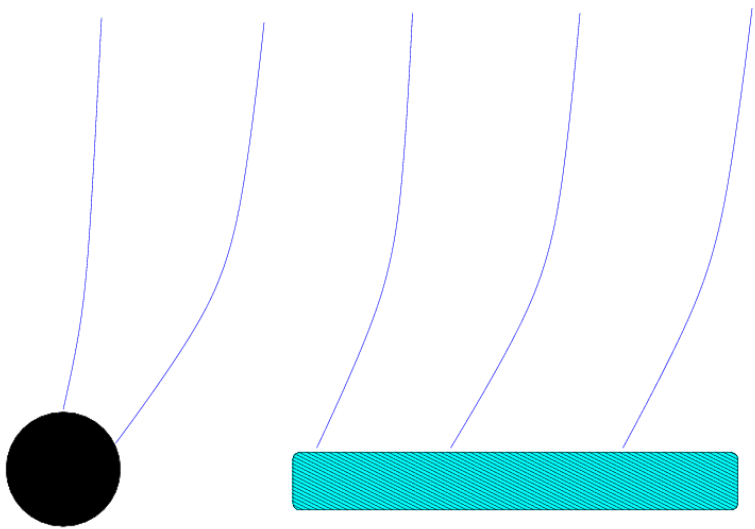}}
\caption{A black hole accretion disk (with the no-flux gap region
boundary condition) threaded by large scale magnetic field that is
parallel to the black hole spin axis far from the accretion
disk. }\label{BlackHole_disk}
\end{figure*}

\begin{figure*}[h!]
\centerline{\includegraphics[angle=-0,scale=0.8]{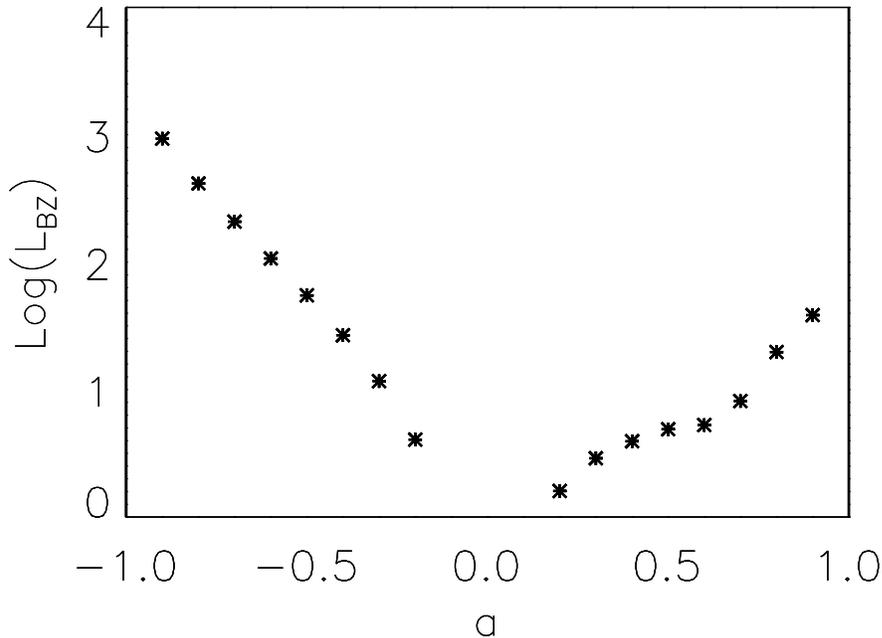}}
\caption{Blandford-Znajek power vs. spin according to
flux-trapping. }\label{Flxvsspin}
\end{figure*}

 The basic feature of our model is illustrated in Figure
 \ref{BlackHole_disk} where magnetic field lines threading the black
 hole are separated from those threading the disk by a gap region (or
 plunge region).  The absence of magnetic field threading the gap
 region is the fundamental assumption of the flux-trapping model
 (Reynolds, et al, 2006).  This assumption has implications for both
 the BZ and BP effects of which the former are illustrated in Figure
 \ref{Flxvsspin}, originating from the numerical solution to the MHD
 equations in a Kerr metric (Garofalo, 2009).  We emphasize the fact
 that maximum BZ power is produced for highly retrograde black hole
 spin, and extend the flux-trapping model to outflows of the BP type,
 with the basic point to motivate the existence of a spin dependence
 in BP power that is also maximized at high retrograde black hole spin
 values.  The model is further described below.

\begin{enumerate}
\item Our accretion disk is described by a Novikov \& Thorne (1974)
disk truncated at the marginally stable orbit, inwards of which is the
gap region.  

\item In the magnetosphere (the region outside of the black hole and
accretion disk) we assume that the plasma density is negligible and
hence that the magnetic field is force free.

\item We assume that no magnetic flux threads the gap or plunge region
of the accretion disk.  Any magnetic flux that is advected inwards
across the radius of marginal stability is immediately added to the
flux bundle threading the black hole.

\item Far away from the black hole and at poloidal angles above the
  accretion disk, we assume the large-scale field is uniform.

\end{enumerate}

%% In this section, we use  the \subsection command to set off
%% a subsection.  \footnote is used to insert a footnote to the text.

%% Observe the use of the LaTeX \label
%% command after the \subsection to give a symbolic KEY to the
%% subsection for cross-referencing in a \ref command.
%% You can use LaTeX's \ref and \label commands to keep track of
%% cross-references to sections, equations, tables, and figures.
%% That way, if you change the order of any elements, LaTeX will
%% automatically renumber them.

%% This section also includes several of the displayed math environments
%% mentioned in the Author Guide.

\section{BP outflows and the cosmological evolution of black hole spin in the flux-trapping model}

\begin{figure*}[h!]
\centerline{\includegraphics[angle=-90,scale=0.4]{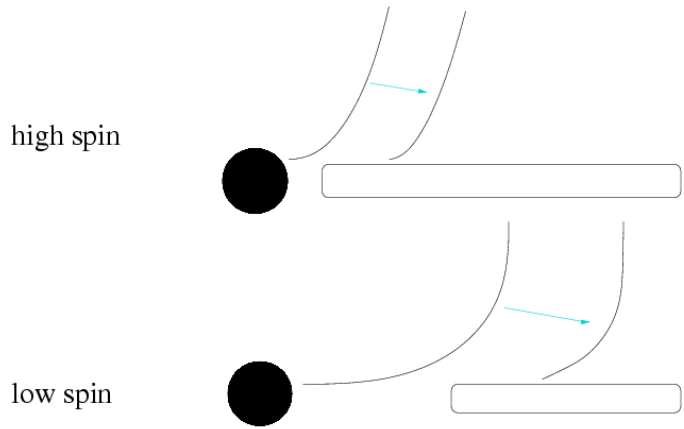}}
\caption{The difference in the geometry between a high spinning black
hole accretion disk (top) and a low spinning or retrograde black hole
accretion disk (bottom). In the high-spin case the gap region is small
and the black hole acquires little magnetic flux from the dragging of
magnetic field through it.  The resulting weak flux bundle on the hole
has little effect on the magnetic field threading the disk whose
geometry remains thus mostly vertical.  In the low-spin or retrograde
case, instead, the gap region is large and it drags a strong flux
bundle onto the horizon.  The strength of this flux bundle compared to
that in the high spin case is such that it is comparatively more
effective in deforming the disk-threading magnetic flux lines which
bend.}\label{spin_geometry}
\end{figure*}

\begin{figure*}[h!]
\centerline{\includegraphics[angle=-0,scale=0.7]{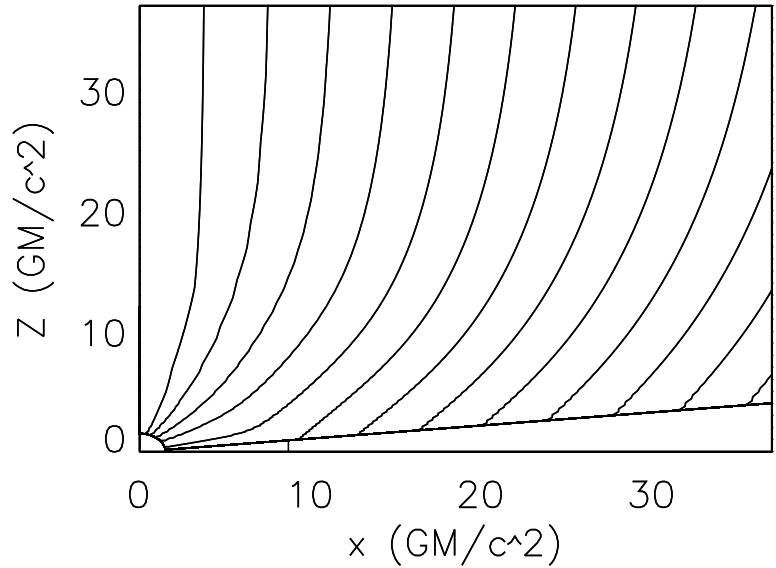}}
\caption{Magnetic configuration for a -0.90 retrograde spinning black
hole and its accretion disk.  Notice the large bend at 30
gravitational radii compared to the high prograde figure.
}\label{spin_negative}
\end{figure*}

\begin{figure*}[h!]
\centerline{\includegraphics[angle=-0,scale=0.7]{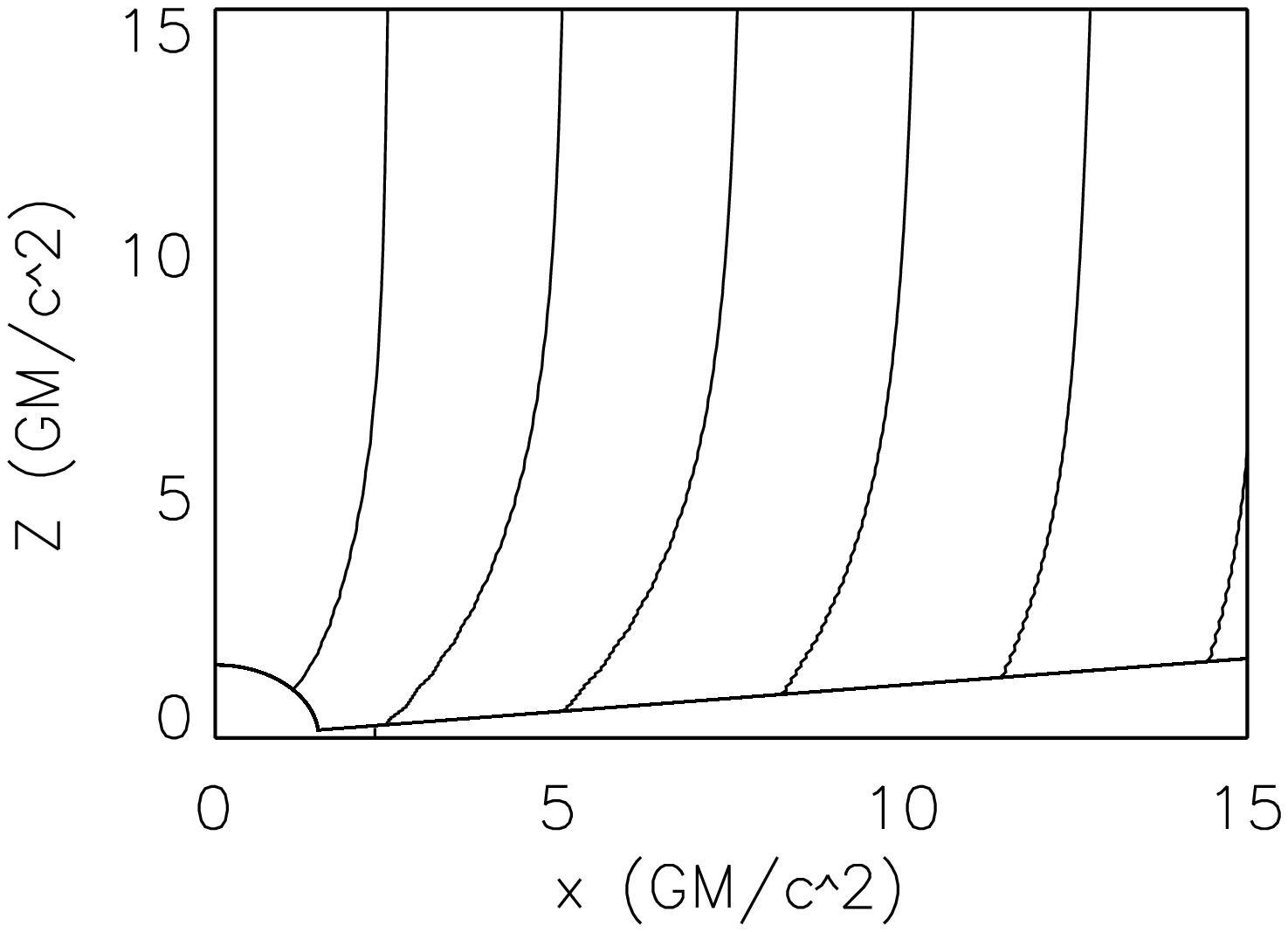}}
\caption{Magnetic configuration for a 0.90 prograde spinning black
hole and its accretion disk.  Notice how the flux lines in the disk
are only slightly bent at 15 gravitational radii unlike the high
retrograde case where they are considerably bent at that
location. }\label{spin_positive}
\end{figure*}

In this section the focus is on the geometry of the magnetic field as
in figure~\ref{BlackHole_disk} and the changes that occur as the spin
of the black hole varies.  Because BP outflows depend on the angle
between the magnetic field and the accretion disk surface, the
emphasis is on how this angle changes with spin.  Despite highlighting
MHD force balance in the force-free magnetosphere, the discussion
remains qualitative, limiting the study to identifying the spin value
for which BP power is maximized.  Magnetic forces between the flux
bundle on the hole and that threading the disk compete at latitudes
above the equatorial plane where the no-flux boundary condition is
imposed (see arrows in Fig.~\ref{spin_geometry}).  Whereas magnetic
pressure/tension of magnetic field lines threading the disk tend to
push the hole-threading flux bundle onto the horizon, the latter
reacts back on the disk-threading magnetic field to limit additional
magnetic field advection onto the black hole.  The bend in the
magnetic field threading the disk stems from the fact that while the
radial inflow of the accreting gas attempts to drag the large scale
magnetic field toward the black hole, the aformentioned magnetic
forces from the flux bundle already threading the black hole, push the
magnetic field lines threading the disk outward.  The greater the
magnetic flux bundle on the black hole, the more effective its ability
to halt additional advection from the disk, and the greater the bend
inflicted on the magnetic field lines threading the disk.  As
Fig.~\ref{spin_geometry} illustrates, the magnitude of the black
hole-threading flux bundle depends on the ability of the gap region
to drag magnetic field inward.  For high prograde-spinning black
holes, the marginally stable circular orbit is close to the black hole
horizon in both coordinate and proper distance which makes the gap
region ineffective at dragging large magnetic flux to the horizon.  In
the low-spin or retrograde case, instead, the inner edge of the
accretion disk at the marginally stable circular orbit is much further
out so the proper distance to the horizon from the disk inner edge is
larger.  This means that slowly spinning or retrograde black holes
acquire magnetic flux via the gap region further out in radial
position compared to their high-spin counterparts, resulting in a
larger magnetic flux bundle on the horizon.  As BP point out, if the
magnetic field lines and the disk surface meet at an angle that is 60
degrees or less, a centrifugally-driven MHD wind is possible.  With
respect to their high-spin counterparts, then, low-spin or retrograde
systems are more likely to exhibit bent magnetic field line
configurations which makes them comparatively better candidates for BP
outflows.  This behavior is seen in the steady-state magnetic field
configurations of the numerical solution (figures \ref{spin_negative}
and \ref{spin_positive} ).  We choose a representative set of disk
parameters such as disk thickness, accretion rate, Prandtl number
etc., and illustrate the geometry of the numerical solution.  We find
that the magnetic field lines threading the retrograde system are bent
well out into the disk.  The high prograde spin system, on the other
hand, displays bent magnetic flux contours only in the innermost
region of the accretion disk and even there the bend is small.  In
short, as the spin of the black hole decreases from high prograde
toward high retrograde values, the magnetic field lines bend
progressively more toward the disk surface.  If we associate this
feature with greater BP outflow power, the arguments suggest that the
spin dependence of BP power in the flux trapping model increases
progressively from high prograde spins to high retrograde spins.
Therefore, like the BZ power, the BP power is maximized for
$a\approx-1$.

Assuming that the BZ and BP mechanisms are the dominant path chosen by
nature to produce outflows and jets in AGN within the context of the
flux-trapping model, leads to the conclusion that retrograde black
hole accretion systems threaded by large-scale magnetic fields, tend
toward prograde black hole spin systems unless some external factors
beyond accretion and black hole rotational energy extraction occur.
Accretion in retrograde systems, in fact, adds angular momentum to the
hole in the prograde sense.  In addition, the BZ power is largest for
high retrograde spins which means that greatest spin energy extraction
occurs to reduce the absolute value of the hole's angular momentum.
The BP mechanism, on the other hand, does not directly affect the
black hole spin.  Clearly, accretion and spin-energy extraction via BZ
both operate to increase the spin away from $a\approx-1$.  How fast it
moves away from $a\approx-1$ and whether it crosses $a=0$ depends on
the rate at which angular momentum is extracted by the BZ mechanism as
well as on the rate at which angular momentum is supplied by
accretion.  The one thing that is clear, though, is that the most
energetic outflows and jets produced in $a\approx-1$ systems are not
stable, so the spin must change.  Thus, if flux-trapping via the
plunge region occurs in nature, the most powerful AGN evolve to lower
energies as their black hole spins become more prograde.

\section{Conclusions}

This work extends the relativistic flux-trapping model to include
outflows of both the BZ and BP type in an effort to constrain the
cosmological evolution of black hole spin (e.g. Brenneman, 2009) and
its possible connection to powerful outflows such as those in radio
loud galaxies (Evans et al., 2009, in press). Our current picture of
the interaction between black holes in AGN and the host galaxy,
suggests a coherent two-way conversation in which the host galaxy
speaks to the black hole about the galaxy via accretion by funneling
matter toward the black hole, and the black hole speaks to the galaxy
about the black hole via outflows that leave signatures of its mass
(Kormendy \& Richstone, 1995; Magorrian et al. 1998; Marconi \& Hunt,
2003; Gebhardt et al. 2000; Ferrarese \& Merritt, 2000; Tremaine et
al, 2002).  If the behavior of the gap region is as fundamental as
assumed here, this two-way conversation includes more, one in which
the more subtle features of the highly non-Newtonian aspects of space
and time that dominate the region close to the center of galaxies are
also revealed.  In fact, the scenario that emerges is one in which the
spin parameter of the central supermassive black hole is not simply
revealed in the generated outflow, but is an active participant in the
dynamics of the latter, to the extent that it sets the scale for the
magnitude of the outflow power.  The overall conclusion of the
assumption that BP and BZ operate within the context of flux-trapping
is that galaxy evolution is tightly coupled to black hole spin.
\label{conclusion}

%\begin{figure*}[h!]
%\centerline{\includegraphics[angle=-90,scale=0.6]{spin_geometry.ps}}
%\caption{The difference in the geometry between a high spinning black
%hole accretion disk (top) and a low spinning or retrograde black hole
%accretion disk (bottom). In the high-spin case the gap region is small
%and the black hole acquires little magnetic flux from the dragging of
%magnetic field through it.  The resulting weak flux bundle on the hole
%has little effect on the magnetic field threading the disk whose
%geometry remains thus mostly vertical.  In the low-spin or retrograde
%case, instead, the gap region is large and it drags a strong flux
%bundle onto the horizon.  The strength of this flux bundle compared to
%that in the high spin case is such that it is comparatively more
%effective in deforming the disk-threading magnetic flux lines which
%bend.}\label{spin_geometry}
%\end{figure*}

%\begin{figure*}[h!]
%\centerline{\includegraphics[angle=-0,scale=0.7]{retrograde_flux.ps}}
%\caption{Magnetic configuration for a -0.90 retrograde spinning black
%hole and its accretion disk.  Notice the large bend at 30
%gravitational radii compared to the high prograde figure.
%}\label{spin_negative}
%\end{figure*}

%\begin{figure*}[h!]
%\centerline{\includegraphics[angle=-0,scale=0.7]{prograde_flux.ps}}
%\caption{Magnetic configuration for a 0.90 prograde spinning black
%hole and its accretion disk.  Notice how the flux lines in the disk
%are only slightly bent at 15 gravitational radii unlike the high
%retrograde case where they are considerably bent at that
%location. }\label{spin_positive}
%\end{figure*}

\section{acknowledgments}

The author thanks David L. Meier for detailed discussion on the BP
effect.  The research described in this paper was carried out at the
Jet Propulsion Laboratory, California Institute of Technology, under a
contract with the National Aeronautics and Space Administration.
D.G. is supported by the NASA Postdoctoral Program at NASA JPL
administered by Oak Ridge Associated Universities through contract
with NASA.  
%Copyright 2009 California Institute of Technology.
%Government sponsorship acknowledged.

\section*{References}

\noindent Bisnovatyi-Kogan, G.S. \& Lovelace, R.V.E., 2007, ApJ, 667, L167

\noindent Bisnovatyi-Kogan, G.S. \& Ruzmaikin, A.A.,1976, Ap\&SS, 42,
401

\noindent Blandford, R. D., \& Payne, D. G. 1982, MNRAS, 199, 883

\noindent  Blandford, R. D., \& Znajek, R. L. 1977, MNRAS, 179, 433 

\noindent  Blandford, R. D., 1976, MNRAS, 176, 465

\noindent  Brenneman, L. Astro2010 Science White Paper

\noindent Evans, D., 2009, ApJ, in press.

\noindent Ferrarese, L., \& Merritt, D., 2000, ApJ, 539, L9

\noindent Garofalo, D., 2009, ApJ, 10, 700   

\noindent Gebhardt, K., et al. 2000, ApJ, 539, L13 

\noindent Kormendy, J., \& Richstone, D., 1995, ARA\&A, 33, 581

\noindent Li, Z-Y, Chiueh, T. \& Begelman M., 1992, ApJ, 394, 459

\noindent Lovelace, R.V.E. 1976, Nature, 262, 649

\noindent Magorrian, J., et al. 1998, AJ, 115, 2285

\noindent Marconi, A., \& Hunt, L.K., 2003, ApJ, 589, L21

\noindent Meier, D. L., 1999, ApJ, 522, 753

\noindent D.L. Meier et al., Science 291 (2001), 84

\noindent Nakamura, M. et al., 2008, ASPC, 386, 373N

\noindent  Reynolds, C. S., Garofalo, D., \& Begelman, M. 2006, ApJ, 651, 1023

\noindent Rothstein D.M. \& Lovelace, R.V.E., 2008, ApJ, 677, 1221

\noindent Tremaine, S., et al 2002, ApJ, 574, 740

\noindent  Vlahakis, N. \& Konigl, A., 2003, ApJ, 596, 1104

\end{document}